\documentstyle[sprocl,graphicx,psfrag]{article}

\def\be{\begin{equation}}
\def\ee{\end{equation}}
\def\bea{\begin{eqnarray}}
\def\eea{\end{eqnarray}}
\newcommand{\eff}{\mbox{\footnotesize\textit{eff\/}}}
\newcommand{\link}{\mbox{\footnotesize\textit{link\/}}}
\newcommand{\plaq}{\mbox{\footnotesize\textit{plaq\/}}}
\newcommand{\copies}{\mbox{\footnotesize\textit{copy\/}}}
\newcommand{\cp}{\mbox{\footnotesize\textit{cp\/}}}
\begin{document}
\title{\uppercase{Center vortices and colour confinement 
\\in lattice QCD}%
\renewcommand{\thefootnote}{\fnsymbol{footnote}}%
\footnote{Plenary talk presented by {\v{S}}.\ Olejn{\'\i}k.
His work was supported in part by the ``Action Austria--Slovakia:
Cooperation in Science and Education'' (Project No.\ 30s12)
and the Slovak Grant Agency for Science (Grant VEGA No.\ 2/7119/2000).} 
\addtocounter{footnote}{-1}}
\author{ROMAN BERTLE, MANFRIED FABER}
\address{Institut f\"ur Kernphysik, Technische Universit\"at Wien,
A--1040 Vienna, Austria
\\E-mail: {\tt bertle@kph.tuwien.ac.at, faber@kph.tuwien.ac.at}}
\author{JEFF GREENSITE}
\address{Physics and Astronomy Department, San Francisco State University,
\\San Francisco, CA 94117, USA, and
\\Theory Group, Lawrence Berkeley National Laboratory,
Berkeley, CA 84720, USA
\\E-mail: {\tt greensit@stars.sfsu.edu}}
\author{\v STEFAN OLEJN{\'I}K}
\address{Institute of Physics, Slovak Academy of Sciences, 
SK--842 28 Bratislava, Slovakia
\\E-mail: {\tt fyziolej@savba.sk}}
\maketitle
\abstracts{We review lattice evidence showing that center-vortex 
condensation is a serious candidate for the mechanism of colour
confinement in quantum chromodynamics.}
%
%
\section{Introduction}\label{intro}
	Any model that contends for the true theory of 
colour confinement in SU($N$) gauge theory has to be able to 
provide an explanation of or an answer to the following facts 
and questions:
\begin{list}{$\bullet$}{\topsep2pt\parsep0pt\itemsep1pt}
\item
All non-zero $N$-ality colour charges are confined.
\item
All zero $N$-ality charges are screened at {\em large\/} distances.
\item
At {\em intermediate\/} distances, potentials between 
higher representation char\-ges rise approximately
linearly, and the corresponding string 
tensions are proportional to the eigenvalues of the quadratic Casimir 
operator of the representation. 
(We dubbed this phenomenon {\em Casimir scaling\/}.)
\item
A deconfinement transition occurs at high temperatures or high 
densities of hadronic matter. What is the nature of the transition? 
What happens to confining configurations at the transition, what 
symmetry gets restored or broken?
\item
What is the relation to other phenomena, like chiral symmetry breaking?
\end{list}
In this talk I will argue that confining configurations that have a chance
to successfully address the above issues, are {\em center vortices\/}.
I will mostly review recent results of our group obtained in lattice
investigations of the confinement problem. Other aspects of the vortex picture
were covered at this Conference in the plenary talk of 
Reinhardt~\cite{Reinhardt:2000nn} 
(see also short contributions of Alexandru~\cite{Alexandru:2000ax}, 
Engelhardt~\cite{Engelhardt:2000xj}, 
and Langfeld~\cite{Langfeld:2000jt}).

	The vortex-condensation model of confinement stems from late
seventies, and was formulated first by `t Hooft~\cite{tHooft:1978hy} 
and developed by many authors.
Let me briefly recall the original idea. To characterize
phases of the gauge theory, Wilson~\cite{Wilson:1974sk} 
suggested to use the loop operator
\be
A(C) =\mbox{Tr}\;\left[ P\;
\exp\left(ig\displaystyle{\oint_C} {\mathbf{A}}_\mu dx^\mu\right)\right].
\ee
In a pure gauge theory, confinement is signalled by 
the leading behaviour of this quantity, namely
\be
W(C)=\langle A(C)\rangle
\sim \left\{ \begin{array}{ccl}
                 e^{-\sigma {\cal{A}}(C)} &\quad\dots\quad&
                                           \mbox{confinement,} \\
                   e^{-\mu {\cal{P}}(C)}  &\quad\dots\quad&
                                           \mbox{deconfinement,} 
                 \end{array} \right.
\ee
where ${\cal{A}}(C)$ is the area of the minimal
surface spanned by the loop, ${\cal{P}}(C)$ its perimeter.
Wilson's loop operator measures the {\em magnetic\/} flux through
$C$ and creates {\em electric\/} flux along $C$.

	`t Hooft~\cite{tHooft:1978hy} suggested another 
operator, measuring the {\em electric\/} flux 
through some other loop $C'$ and creating {\em magnetic\/} flux along $C'$.
It was defined through commutation relation with $A(C)$ 
($n$ is the {\em linking number\/} between $C$ and $C'$):
\be
A(C)B(C') = B(C')A(C)
\exp\left(2\pi i n/N\right).
\ee
The expected behaviour of `t Hooft's operator is opposite to the Wilson
loop:
\be
\langle B(C')\rangle
\sim \left\{ \begin{array}{ccl}
     e^{-\mu {\cal{P}}(C')}   &\quad\dots\quad&\mbox{confinement,} \\
     e^{-\sigma {\cal{A}}(C')}&\quad\dots\quad&\mbox{deconfinement.} 
                 \end{array} \right.
\ee
The effect of the `t Hooft loop operator $B(C')$ on physical fields
is a gauge transformation which is {\em singular\/} on the curve $C'$.
It creates a thin vortex of magnetic flux (line-like in 3D,
surface-like in 4D). To avoid infinite energy, the singularity
along the loop $C'$ has to be smoothened: vortices acquire a {\em core\/}
of certain thickness. {\em Many things in this talk will be related
in some way to the thickness of the vortex core.\/}

	The essence of the vortex model of confinement is that 
the QCD vacuum is filled with closed magnetic vortices that have 
the topology of tubes (in 3D) or surfaces (in 4D). Thick vortices 
condense in the vacuum.
A simple argument can explain the area law for the Wilson loop through
fluctuations in the number of vortices linked to the loop.    

	It is a sort of paradox that this idea has for years not been
subjected to extensive numerical tests in lattice simulations.
However, in the last couple of years the idea returned to the stage
and got into the focus of attention. This turn was due to the
discovery of center dominance in maximal center 
gauge~\cite{DelDebbio:1997mh,DelDebbio:1998uu}, and
the following accumulation of evidence in favour of the
picture~\cite{deForcrand:1999ms}$^{-}$\cite{Bakker:1999qh}.

	The outline of my talk is the following: In 
Section \ref{strong_coupling} I will briefly sketch our recent 
strong-coupling calculation~\cite{Faber:2000um}, which shows
that center vortices are stable saddlepoint configurations of a long-range
effective action that can be derived from the usual lattice Wilson action.
Then (Section \ref{MCG}), I will introduce our method to detect center
vortices that is based on center projection in maximal center gauge, 
and review some interesting results obtained in numerical
simulations. Section \ref{copies}
then addresses a question, raised recently by 
Bornyakov et al.~\cite{Bornyakov:2000cd}, 
how physical results depend on the number of gauge copies used
in iterative MCG fixing.
Finally, I will argue in Section \ref{Casimir_scaling} that the observed 
Casimir scaling can be accommodated with the vortex picture, and that
the thickness of the vortex core is vital for the explanation
of the scaling behaviour.   

%
\section{Confining Configurations of the Long-Range Effective
Action}\label{strong_coupling}
	There are various ways to argue for the center of the gauge group 
being important for the confinement mechanism. The asymptotic
string tension for charges in the representation $r$ depends only
on its $N$-ality (i.e.\ $Z_N$ charge); the deconfinement transition is
associated with spontaneous breaking of the global $Z_N$ symmetry;
etc. We have recently provided another argument: if one calculates 
an effective action at larger distances (and strong coupling)
starting from the usual Wilson action, its stable saddlepoints are 
thick center vortices.

\begin{figure}
\centering
\psfrag{U}{$\scriptstyle\tilde{U}$}\psfrag{V}{$\scriptstyle V$}
\includegraphics[width=0.6\textwidth]{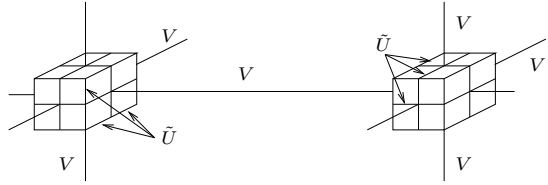}
\caption{Unintegrated links on the coarse $V$-lattice.}
\label{v00fig1}
\end{figure}

	The calculation proceeds in the following 
way~\cite{Faber:2000um}: We start from the
pure gauge theory at strong coupling on a fine lattice 
(links on the fine lattice are 
denoted generically as $U$), and try to calculate the effective
action on a coarse lattice with links $V$, where the coarse lattice spacing
is $L$ times the original $U$-lattice spacing. The idea is to derive an 
effective action where the {\em leading\/} contributions
to any Wilson loop on the $V$-lattice are obtained from a {\em local\/}
action. To achieve this, we integrate over all links on the $U$-lattice
except $\tilde{U}$-links on small 2-cubes around sites of the 
$V$-lattice, as shown in Fig.\ \ref{v00fig1}.%
\footnote{We work in 3 dimensions. The extension to 4D should be 
straightforward.} 

	This defines effective action $\tilde{S}_L$:
\bea
\nonumber
Z&=&\int [DV] \int \prod_{l\in\atop\mbox{\tiny 2-cubes}} [d\tilde{U}_l]
\left\{\int \prod_{l''\not\in\atop\mbox{\tiny 2-cubes}} [dU_{l''}]
\prod_{l'} \delta\left[V^\dagger_{l'}(UU..U)_{l'} -I\right] 
e^{S_W[U]}\right\}\\ 
&\phantom{=}&\qquad\qquad
\equiv\int [DV] \int \prod_{l\in \mbox{\small 2-cubes}} [d\tilde{U}_l]\;
\exp\left(\tilde{S}_L[V,\tilde{U}]\right).
\eea

\begin{figure}[b!]
\centering
\includegraphics[width=0.6\textwidth]{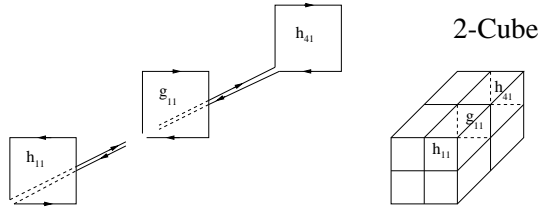}
\caption{Plaquette variables on the 2-cubes.}
\label{v00fig2}
\end{figure}

	Instead of $\tilde{U}_l$ we introduce in the 2-cubes group-valued
plaquette variables $h,g$ (Fig.\ \ref{v00fig2})~\cite{Batrouni:1982dx}. 
After integrating out the $g$ variables,  the result is 
(displaying only terms of low order in both $h$ and~$\beta$):
\bea
\nonumber
Z&\approx&\int [DV] [Dh] \prod_{K} 
\Bigg\{ 1 + 2 \left({\beta\over 4}\right)^3 \sum_{c\in K} 
\chi_{1\over 2}[(hhh)_c]\\
\nonumber
&&\phantom{bla}\qquad\qquad\qquad\qquad
2 \left({\beta\over 4}\right)^4\sum_{\mbox{\small adj. }c_1c_2\in K} 
\chi_{1\over 2}[(hhh)_{c_1}(hhh)_{c_2}] + \dots \Bigg\}\\
\nonumber
&\times&\exp\left[{\beta\over 2} \sum \mbox{Tr}[h] 
    + 2\left(\beta\over 4\right)^{4(L-2)}\sum_{l'}
  f_{l'}^{ijkl} \mbox{Tr}[h^\dagger_{ij}V_{l'}
            h^\dagger_{kl}V^\dagger_{l'}] \right.\\
&&\phantom{bla}\qquad\qquad\qquad\qquad
\left. + 2\left({\beta\over 4}\right)^{L^2} \sum_{P'}
              \mbox{Tr}[V V V^\dagger  V^\dagger] \right].
\eea

	This resembles an adjoint-Higgs Lagrangian, with an SU(2)
gauge field $V_\mu$ coupled to 24 ``matter'' fields $h$ in the
adjoint representation.  Note that for large {$L$}, 
the ``Higgs'' potential term is much larger than the ``kinetic''
and pure-gauge
({$V$}-plaquette) terms, so the {$h$}-fields 
fluctuate almost independent of {$V_\mu$}.

	We then make a unitary gauge fixing of the $h$-fields
(breaking {SU(2) to $Z_2$}), integrate out the remaining
$h$ d.o.f., and arrive at:
\bea
\nonumber
S_{\eff}[V]&\approx&
S_{\link}[V,\langle h \rangle_h] + S_{\plaq}[V]\\
\nonumber
&=&2\left({\beta\over 4}\right)^{4(L-2)}
\sum_{l'} f_{l'}^{ijkl}\mbox{Tr} \left[
      \langle h^\dagger_{ij} \rangle_h V_{l'} 
      \langle h^\dagger_{kl} \rangle_h V^\dagger_{l'} \right]\\
&+& 2\left({\beta\over 4}\right)^{L^2}
             \sum_{P'} \mbox{Tr}[VVV^\dagger V^\dagger].
\eea

	Let us now look for saddlepoints.  $S_{\link}$ is maximized by
\be
V_\mu(\vec{n})=Z_\mu(\vec{n}) 
                       \times g(\vec{n}) g^\dagger(\vec{n}+\mu),
\qquad\qquad
Z_\mu =\pm \mathbf{1},
\ee
where $g(\vec{n}) g^\dagger(\vec{n}+\mu)$ is fixed by the particular 
unitary gauge choice, while $S_{\plaq}$ is maximized if 
$ZZZZ=+\mathbf{1}$.  This is the unitary gauge {\em ground state\/}.

	One can create a {\em thin center vortex\/} on this state 
by a discontinuous gauge transformation, e.g.\
\be
Z_y(\vec{n}) =\left\{ \begin{array}{rcl}
                    -\mathbf{1} & \quad\dots\quad &n_1 \ge 2, ~ n_2=1, \cr 
                    +\mathbf{1} & \quad\dots\quad &\mbox{otherwise,} 
                      \end{array} \right.\qquad
Z_x(\vec{n}) = Z_z(\vec{n}) = \mathbf{1}.
\ee         
This configuration is stationary:  $S_{\link}$ is
still a maximum (insensitive to center), $S_{\plaq}$ is extremal 
(maximal or minimal) on all plaquettes.

	Stability depends on the eigenvalues of
\be
     \frac{\delta^2 S_{\eff}}{\delta V_\mu(n_1) \delta V_\nu(n_2)} =
        \frac{\delta^2 S_{\link}}{\delta V_\mu(n_1) \delta V_\nu(n_2)}
      + \frac{\delta^2 S_{\plaq}}{\delta V_\mu(n_1) \delta V_\nu(n_2)}, 
\ee
where
\be
\frac{\delta^2 S_{\link}}{\delta V\delta V}\sim 
\left({\beta\over 4}\right)^{4(L-2)+12},\qquad\qquad
\frac{\delta^2 S_{\plaq}}{\delta V\delta V}         
\sim\left({\beta\over 4}\right)^{L^2}.
\ee

	Crucial fact is now the following: for $\beta/4 \ll 1$ and
\be
      4(L-2) + 12 < L^2  \qquad\Longrightarrow\qquad L \ge 5
\ee
the contribution of $\delta^2 S_{\plaq}/\delta V \delta V$ 
to the stability matrix (and therefore to its eigenvalues) 
is negligible compared to $\delta^2 S_{\link}/\delta V \delta V$.

	This has (at least) three important implications:
\begin{list}{$\bullet$}{\topsep2pt\parsep0pt\itemsep1pt}
\item{\em Vortex stability:\/}
The thin vortex is a stable saddlepoint of the 
full effective action $S_{\eff}$ at {$L\ge 5$}.
\item{\em Vortex thickness:\/}
A ``thin'' vortex on the $V$-lattice means thickness $< L$ on the
$U$-lattice.  This means that stable center vortices are $\approx 4-5$
lattice spacings thick.
This is the distance where the adjoint string breaks at strong couplings! 
\item{\em Percolation:\/}
From $S_{\eff}$, we see that center vortices in $D=3$ 
cost action
$8 \left( {\beta/4} \right)^{L^2} / \mbox{unit length}$,
while the entropy is $O(1)/ \mbox{unit length}.$

Entropy $\gg$ action implies that vortices percolate through the
lattice, and confine nonzero $N$-ality charges.
\end{list}

	However, the above result has been obtained in the strong-coupling 
approximation. To study the role of vortices at weak couplings, we resort 
to numerical simulations.

%
\section{Center Projection in Maximal Center Gauge}\label{MCG}
	The model of `t Hooft relies on no particular gauge.
However, without gauge fixing, one can derive
certain relations that, at the first sight, seem related to vortices,
but apparently have no physical relevance~\cite{Faber:1999en}.
Gauge fixing turns out to be also useful in pin-pointing
the relevant (for vortices and confinement) degrees of freedom.
One can then study in detail
e.g.\ their topological properties~\cite{Bertle:1999tw}, 
behaviour across deconfinement
phase transition~\cite{Engelhardt:2000fd}, etc.
 
	We proposed to identify vortices in thermalized lattice 
configurations in the following steps~\cite{DelDebbio:1997mh,DelDebbio:1998uu}:
\begin{list}{$\bullet$}{\topsep2pt\parsep0pt\itemsep1pt}
\item
First we fix to {\em maximal center gauge\/}%
\footnote{For the continuum counterpart of MCG see the talk of
Reinhardt~\cite{Reinhardt:2000nn,Engelhardt:2000xw}.} 
by maximizing the expression:
\be\label{MCG_condition} 
{\cal{R}}\equiv\sum_{x,\mu}\;\Bigl| \mbox{Tr}[U_\mu(x)] \Bigr|^2\;.
\ee
This in fact is {\em adjoint Landau gauge\/}; the above 
condition is equivalent to maximizing
\be
{\cal{R}}'\equiv\sum_{x,\mu}\;\mbox{Tr}[U^A_\mu(x)]\;.
\ee
\item
Then we make {\em center projection\/} by replacing:
\be
U_\mu(x)\rightarrow Z_\mu(x) \equiv \mbox{sign Tr}[U_\mu(x)]\;.
\ee
\item
Finally we identify excitations ({\em P-vortices\/}) of the resulting
$Z_2$ lattice configurations.
\end{list}

	A series of findings shows that vortices identified via 
center projection in MCG, are crucial for colour confinement 
and related phenomena:
\begin{list}{$\bullet$}{\topsep2pt\parsep0pt\itemsep1pt}
\item
{\em Center dominance:\/} It was observed both in 
SU(2)~\cite{DelDebbio:1997mh,DelDebbio:1998uu} and (less convincingly) 
in SU(3) lattice gauge theory~\cite{Faber:2000sq} that the string tension
obtained from center-projected configurations in MCG 
(or Laplacian center gauge~\cite{deForcrand:2000pg}) 
agrees remarkably well with the asymptotic string tension of the full theory.
\item
{\em P-vortices locate center vortices:\/}~\cite{DelDebbio:1998uu} 
One can define ``vortex-limited''
Wilson loops $W_n(C)$, i.e.\ Wilson loops evaluated on the original,
unprojected lattice, on a subensemble of configurations in which
$n$ P-vortices pierce the minimal area of the loop. One expects
$W_n(C)/W_0(C)$ to approach $(-1)^n$ asymptotically; this expectation
is confirmed for large enough loops in numerical simulations.
\item
{\em P-vortices locate physical 
objects:\/}~\cite{Langfeld:1998jx,DelDebbio:1998uu} Vortex density scales
according to the asymptotic freedom, like a dimensionful quantity 
$\sim \Lambda^2$.
\item
{\em Center vortices are correlated with 
confinement:\/}~\cite{DelDebbio:1997mh,deForcrand:1999ms}
Removal of center vortices destroys confinement.
\item
{\em Center vortices are correlated with chiral symmetry
breaking:\/}~\cite{deForcrand:1999ms}
Removal of center vortices restores chiral symmetry.
\item
{\em Center vortices cause non-trivial topology:\/}~\cite{deForcrand:1999ms}
Lattice configurations without vortices belong to the
{\em trivial topological sector\/} (have zero topological 
charge).
\item
{\em Deconfinement can be understood 
as a center vortex percolation 
transition:\/}~\cite{Engelhardt:2000fd,Bertle:1999tw}
In the confined phase, most P-plaquettes belong to a single huge
vortex cluster percolating through the whole lattice volume.
In the deconfined phase, vortex clusters cease percolating
in {\em space\/} slices, while they continue percolating in {\em time\/}
slices. Most vortices are short loops, winding in the time direction.
In this way, perimeter-law behaviour of time-like Wilson loops 
can peacefully coexist with area-law behaviour of space-like
Wilson loops.
\end{list}
%

%
\section{Center Dominance, Gauge Copies, and Lattice Size}
\label{copies}
	Though the above list of indications in favour of our vortex 
identification procedure looks quite impressive, a word of caution 
is necessary. MCG, similar to Landau or maximal abelian gauge, suffers from 
the {\em Gribov copy problem\/}. The iterative procedure, used for gauge 
fixing, converges to a local maximum which is slightly different 
for every gauge copy of a given lattice configuration.
At the first sight, Gribov copies in MCG do not seem to be a severe
problem in our procedure; it appears that P-vortex locations vary 
comparatively little, from copy to copy~\cite{DelDebbio:1998uu}.  

	However, a serious drawback might be a strong dependence
of physical  results (Creutz ratios, vortex density, etc.)
on the number of gauge copies used in MCG maximization procedure.
For each lattice configuration one can generate a set of its random copies, 
fix all these copies to MCG, and evaluate quantities of interest on the
``best'' copy, that is the one with the highest value of ${\cal R}$
in Eq.\ (\ref{MCG_condition}). Such an investigation has recently 
been reported by Bornyakov et al.~\cite{Bornyakov:2000cd}  
They e.g.\ show that 
the $N_{\copies}\to\infty$ extrapolation of the center-projected
string tension underestimates the full string tension by as much as 
30\% (at $\beta=2.5$)! This seems to indicate that the observed
center dominance in MCG (obtained from a small number of copies)
might have been just a numerical coincidence.
\begin{figure}[t!]
\centering
\includegraphics[width=0.6\textwidth]{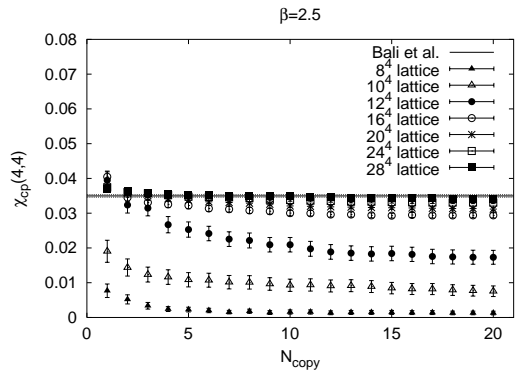}
\caption{The dependence of the Creutz ratio $\chi (4,4)$
on the number of gauge
copies $N_{\copies}$ for $\beta=2.5$, for various lattice sizes.}
\label{v00fig3}
\end{figure}

\begin{figure}[t!]
\centering
\includegraphics[width=0.6\textwidth]{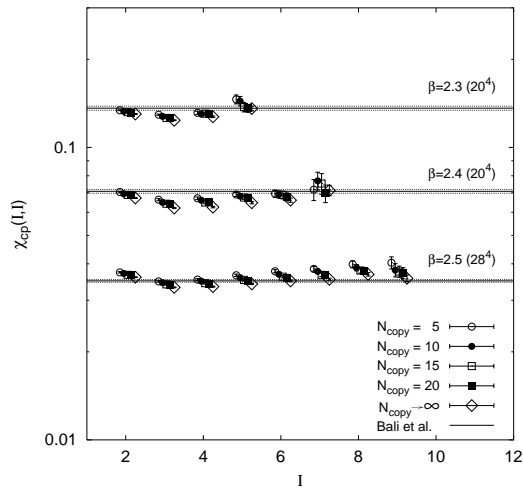}
\caption{Center-projected Creutz ratios $\chi_{\cp}(I,I)$
vs.\ $I$, from our largest lattices.}
\label{v00fig4}
\end{figure}

	In our opinion, the reported strong disagreement is due to finite
size effects. The lattices used by Bornyakov et al.~\cite{Bornyakov:2000cd}
seem rather small compared to those used in our earlier studies. In fact,
what is ``rather small'' or ``large enough'' depends on what are the
typical sizes of vortex cores. Independent sources of estimating
vortex thickness~\cite{DelDebbio:1998uu,deForcrand:2000kr,Kovacs:2000sy}
point towards a value of a little over one fermi,
i.e.\ center vortices are $\approx$ 12--14 lattice spacings thick
at $\beta=2.5$ and the lattice size, used by Bornyakov et al. ($16^4$),
seems too small. We have therefore repeated their analysis
on lattices of various sizes; it turns out that on sufficiently
large lattices the dependence on the number of gauge copies is
much weaker, and center dominance is quite accurate.
This fact is illustrated in Figs.\ \ref{v00fig3} and 
\ref{v00fig4}, details can
be found in our recent publication~\cite{Bertle:2000qv}.%
\footnote{This might not be the final answer to the problem.
We have recently been informed that, using a combination of simulated
annealing and the usual (over)relaxation, one can still
find copies with higher maxima of ${\cal R}$ 
[V.\ Bornyakov, M.\ Polikarpov, private communication].
Also, using Landau gauge preconditioning, one reaches a higher maximum,
but loses center dominance~\cite{Kovacs:1999st}. 
Though we were able to find a cause of 
this problem, the loss of what we called 
``vortex-finding property''~\cite{Faber:1999gu} 
during LG fixing, it is clear that a better gauge-fixing procedure 
and/or a modification of the gauge-fixing condition, is needed. 
Some alternatives have been 
proposed~\cite{Alexandrou:2000iy,Bakker:2000zz,Langfeld:2000jt}.}
%

%
\section{Casimir Scaling}
\label{Casimir_scaling}
	We have repeatedly emphasized that trying to understand the 
confinement mechanism one has to be able to explain the problem of 
Casimir scaling. If we look at string tensions for higher representation 
colour charges, we find different behaviour in two distinguishable regimes:
\begin{list}{$\bullet$}{\topsep2pt\parsep0pt\itemsep1pt}
\item
{\em Casimir-scaling regime:\/} At intermediate distances, from the onset
of confinement to the onset of screening, the string tension is 
roughly proportional the quadratic Casimir. For SU(2), e.g.,
\be
{\sigma_j\over\sigma_{1/2}} \approx {4\over 3} j(j+1).
\ee
\item
{\em Asymptotic regime:\/} The SU($N$) quark colour 
charge is eventually screen\-ed by gluons to the lowest representation with 
the same transformation properties under $Z_N$.  For 
SU(2):
\be
\sigma_j = \left\{ \begin{array}{ccl}
         \sigma_{1/2} & \quad\dots\quad & j=\mbox{half-integer,} \cr
         0    &  \quad\dots\quad & j=\mbox{integer.} \end{array} \right.
\ee
\end{list}
As $N\rightarrow \infty$, colour screening is suppressed, Casimir scaling
is exact.  But even at $N=2,\ 3$ there is a Casimir-scaling
region of finite extent~\cite{Bali:2000un,Deldar:2000vi}. 

	Center vortices are good at explaining 
the asymptotic region.  Loops which transform trivially under the 
group center (such as the adjoint) are unaffected by fluctuations 
in the center elements, and therefore shouldn't get an area law. 
This is strictly true, however, only  for {\em thin\/} center vortices 
(P-vortices).

	Is there a way of reconciling approximate Casimir scaling at 
intermediate distances with center vortices? The answer is apparently yes, 
and the crucial ingredient in the explanation of Casimir scaling through 
vortices is the vortex thickness. 
If the slice of a thick vortex, in the plane of a Wilson loop, is entirely
inside (or entirely outside) the loop, the effect is the same as in
the thin vortex model. However, if the vortex overlaps the 
loop perimeter, one can envisage its effect on the loop 
by insertion of a group element $G$, 
interpolating smoothly between $-\mathbf{1}$ and $\mathbf{1}$ (as
would be the case in an abelian theory). Were the vortices relatively
thin on average, this effect would be unimportant; it would lead
to perimeter-like correction to Wilson loops up to sizes comparable 
to the size of the vortex core. But, as already discussed in the previous
section, there are good reasons to believe that vortices are quite 
thick. This enables one to explain the gross features of the
behaviour of string tensions both in the Casimir-scaling and 
asymptotic regions. A simple model has been worked out in our earlier 
work~\cite{Faber:1998rp} (see also~\cite{Deldar:1999yy}).

	Bali~\cite{Bali:2000un} and Deldar~\cite{Deldar:2000vi} 
have recently published results of high statistics
computations of higher representation potentials in the SU(3) lattice gauge 
theory. Casimir scaling turns out to be quite precise, much more precise than
in our over-simplified model. We believe this is not a real problem 
for center vortices, and our model and Ansatz for vortex-core profiles
can be refined to explain the observed behaviour. 

	Finally I want to mention that to be 
compatible with $N\rightarrow\infty$ where Casimir
scaling is exact, the vortex thickness should grow with $N$.
Some evidence for such a growth has recently been presented
by Montero~\cite{Montero:2000pb}.

%
\section{Epilogue}
\label{conclusion}
\begin{figure}[t!]
\centering
\begin{tabular}{p{0.42\textwidth}p{0.42\textwidth}}
\includegraphics[height=0.29\textwidth]{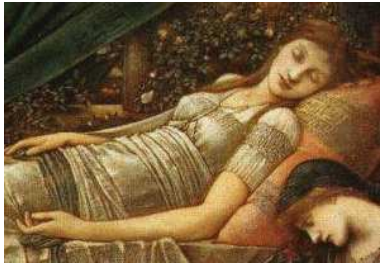}&
\includegraphics[height=0.29\textwidth]{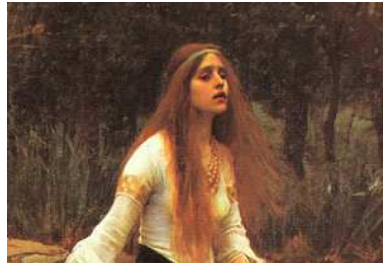}\\
\footnotesize (a) E.\ Burne--Jones (1833--1898):
{\em The Sleeping Princess (Briar Rose)\/}, detail.&
\footnotesize (b) J.~W.~Waterhouse (1849--1917): 
{\em The Lady of Shalott\/}, detail.
\end{tabular}
\caption{``Confinement I'' (1994) 
versus ``Confinement IV'' (2000).}
\label{v00fig5}
\end{figure}

	The first in this series of Conferences took place in 1994 at the 
Lake of Como. There is no mention of the vortex model of confinement in
its Proceedings. The idea, after its happy infant years, was sleeping
deeply at that time like the Briar Rose (Fig.~\ref{v00fig5}a). I tried to
convince you in my talk that now, six years later,
the Sleeping Beauty is up after a long sleep, and looks very 
attractive (Fig.~\ref{v00fig5}b). Some problems, however, persist.
I sincerely hope that, in the years to come, she will not fall asleep
again, or turn into the Maleficent Queen.

%
\section*{Acknowledgments}
	I am grateful to Wolfgang Lucha, Nora Brambilla, Khin Maung Maung 
and the other organizers for inviting  me to present the talk at this 
stimulating meeting.

%

%
\section*{References}

\end{document}